\documentclass[10pt,conference]{IEEEtran}
\usepackage{comment}
\usepackage[T1]{fontenc}
\usepackage[utf8]{inputenc}
\usepackage[english]{babel}
\usepackage[style=ieee, maxnames=9, minnames=9, date=year, doi=false, isbn=false, url=false, backend=biber, sortcites]{biblatex}
\usepackage{nicefrac}
\usepackage{mathtools,amssymb,amsfonts, amsthm}
\usepackage{amsmath}
\usepackage{stmaryrd}
\usepackage{csquotes}
\usepackage{graphicx}
\usepackage{subcaption}
\usepackage{multirow, booktabs}
\usepackage[binary-units,detect-all]{siunitx}
\usepackage[usenames,dvipsnames]{xcolor}
\usepackage{tikz}
\usepackage{subcaption}
\usepackage{soul}
\usetikzlibrary{external}
\usetikzlibrary{positioning}
\usepackage[textsize=tiny]{todonotes}
\usepackage{microtype}
\usepackage{algorithm,algpseudocode}
\usepackage{hyperref}
\usetikzlibrary{calc, positioning, fit, decorations.pathreplacing}
\usepackage{flushend}
\usepackage{float}
\usepackage{yquant}

\usepackage{microtype}
\usepackage{listings}
\usepackage{xcolor}

\definecolor{codegreen}{rgb}{0,0.6,0}
\definecolor{codegray}{rgb}{0.5,0.5,0.5}
\definecolor{codepurple}{rgb}{0.58,0,0.82}
\definecolor{backcolour}{rgb}{0.95,0.95,0.92}

\lstdefinestyle{mystyle}{
	backgroundcolor=\color{backcolour},   
	commentstyle=\color{codegreen},
	keywordstyle=\color{magenta},
	numberstyle=\tiny\color{codegray},
	stringstyle=\color{codepurple},
	basicstyle=\ttfamily\footnotesize,
	breakatwhitespace=false,         
	breaklines=true,                 
	captionpos=b,                    
	keepspaces=true,                 
	numbers=left,                    
	numbersep=5pt,                  
	showspaces=false,                
	showstringspaces=false,
	showtabs=false,                  
	tabsize=2,
}

\makeatletter
\let\oldtheequation\theequation
\renewcommand\tagform@[1]{\maketag@@@{\ignorespaces#1\unskip\@@italiccorr}}
\renewcommand\theequation{(\oldtheequation)}
\makeatother

\newtheorem{example}{Example}

\AtBeginDocument{}

\hypersetup{
	pdftitle={Title},
	pdfsubject={International Symposium on Multiple-Valued Logic, 2024},
	pdfauthor={Jagatheesan Kunasaikaran, Kevin Mato, and Robert Wille}
}

\renewcommand{\baselinestretch}{.796}

\begin{document}
	\title{A Framework for the Design and Realization of\\Alternative Superconducting Quantum Architectures}
	\title{A Framework for the Design and Realization of\\Alternative Superconducting Quantum Architectures}
	
	\author{
		\IEEEauthorblockN{%
			Jagatheesan Kunasaikaran\IEEEauthorrefmark{1}\hspace{1cm}%
			Kevin Mato\IEEEauthorrefmark{1}\hspace{1cm}%
			Robert Wille\IEEEauthorrefmark{1}\IEEEauthorrefmark{4}%
		}
		\IEEEauthorblockA{\IEEEauthorrefmark{1}Chair for Design Automation, Technical University of Munich, Munich, Germany}
		\IEEEauthorblockA{\IEEEauthorrefmark{4}Software Competence Center Hagenberg (SCCH) GmbH, Hagenberg, Austria}
		\IEEEauthorblockA{
			\href{mailto:jagatheesan.kunasaikaran@tum.de}{jagatheesan.kunasaikaran@tum.de},
			\href{mailto:kevin.mato@tum.de}{kevin.mato@tum.de},
			\href{mailto:robert.wille@tum.de}{robert.wille@tum.de}}%
		\IEEEauthorblockA{\url{https://www.cda.cit.tum.de/research/quantum/}\vspace{-.4cm}}
	}
	
	\maketitle
	\begin{abstract}
		Superconducting quantum hardware architectures have been designed by considering the physical constraints of the underlying physics. These general-purpose architectures leave room for customization and optimization that can be exploited with alternative architectures specific to the quantum applications that will be executed on the quantum hardware. However, the corresponding design steps are hardly integrated yet and still heavily relies on manual labor. In this work, we provide an initial software framework that aims at providing a foundation to address this drawback. To this end, we first review the design of superconducting quantum hardware architectures and, afterwards, propose a cohesive framework encapsulating the design flow of an application-specific quantum hardware architecture. The resulting framework integrates high-level architecture generation optimized for a quantum application, the physical layout of the architecture, as well as optimization of the layout in a methodical manner. The framework with a reference implementation is available as part of the \emph{Munich Quantum Toolkit}~(MQT) at \href{https://github.com/cda-tum/dasqa}{https://github.com/cda-tum/dasqa} under an \mbox{open-source} license.
	\end{abstract}
	\section{Introduction}
	
	Quantum computers have been shown to offer substantial speedups over classical computers in solving specific problems. Notable examples of such problems include algorithms found in fields such as chemistry \cite{McArdleQuantum2020}, machine learning \cite{JacobQML2017}, biology~\cite{CordierQuantumBiology2022}, finance~\cite{EggerQuantumFinance2020}, and more~\cite{quetschlich2023towards}.
	
	To realize this promised speedup over classical computers, various quantum hardware architectures such as superconducting~\cite{Krantz2019}, trapped ions~\cite{HartmutIon2008, schoenberger2023using}, or ultracold atoms~\cite{JakschColdAtom2005} have been proposed among others. Particularly, superconducting architectures have been in the forefront of quantum hardware innovation due to their ability to encode quantum information in macroscopic circuit components~\cite{Krantz2019,RasmussenSuperconducting2021}.
	
	However, thus far, the design of these superconducting quantum hardware architectures has mostly been driven with physical constraints in mind. This frequently led to architectures where possible interactions between qubits are limited and corresponding mapping methods are needed (as part of compilers/transpilers) to realize an application on a superconducting architecture (see, e.g.,~\cite{LiTackling2019,Cowtan2019,WilleMapping2019,Tan2020}). They usually require the addition of so-called SWAP gates to realize the needed qubit interactions---severely decreasing the fidelity of the considered application.
	
	While this obviously cannot be avoided completely, recent approaches suggested alternative superconducting quantum architectures that aim to satisfy physical constraints in an application-specific fashion (see, e.g.,~\cite{Li2019,Deb2020,Lin2022}). But tool support to properly design \emph{and} realize them in an automated and integrated fashion hardly exists yet---leaving the designers rather alone with a design task requiring expertise from several domains (such as architecture design and physical design). 
	
	In this paper, we present an initial software framework as well as a reference implementation which aims at supporting the designer in this task. The framework covers the needed steps in an automated and integrated fashion and, hence, allows designers to design and realize alternative and application-specific superconducting quantum architectures in a push-button manner. Furthermore, the framework is realized in a modular and extendable fashion---allowing designers to easily extend it with further methods and, hence, to easily experiment with different design choices. 

	In the remainder of this paper, details of the framework are provided. To this end, we first review superconducting quantum computing architectures and their physical constraints in \autoref{sec:superconducting_quantum_computing_architecture_and_coupling_constraints} and describe the needed design flow towards alternative architectures in \autoref{sec:design_and_realization_of_alternative_architectures}. This also includes the motivation for the proposed framework, which is presented in detail in  \autoref{sec:proposed_framework}---including an overview and details on the implementation as well as descriptions on how to use and how to extend it. Finally, \autoref{sec:conclusion} concludes the paper.
	The proposed framework is publicly available as part of the \emph{Munich Quantum Toolkit}~(MQT) at \href{https://github.com/cda-tum/dasqa}{https://github.com/cda-tum/dasqa}. 
	
	\section{Superconducting Quantum Computing Architectures and Coupling Constraints}
	\label{sec:superconducting_quantum_computing_architecture_and_coupling_constraints}
	
	In order to keep this work self-contained, this section briefly reviews existing superconducting quantum computing architectures and the corresponding coupling constraints. Based on this, it also reviews one of the most challenging tasks in the compilation of quantum circuits, which is caused by these constraints.
	
	\subsection{Superconducting Quantum Computing Architectures}
	
	A prominent technology to construct multi-qubit quantum processors is superconducting~\cite{Krantz2019}. Compared to other technologies (i.e, trapped ions~\cite{HartmutIon2008, schoenberger2023using},  ultracold atoms~\cite{JakschColdAtom2005}) which encodes the quantum information in microscopic systems, superconducting is macroscopic in size. This macroscopic nature allows the manipulation of energy levels, coupling strengths, and coherence rates in the circuit through the configurations of macroscopic circuit elements---opening a vast design space for the quantum hardware designer~\cite{Krantz2019, RasmussenSuperconducting2021}.
	
	Taking note of the advantages of superconducting quantum architectures, industrial participation in realizing these architectures has been accelerating with the public access of superconducting-based quantum hardware provided, e.g., by IBM Q,~\cite{IBMCompute2021}, Rigetti, and OQC quantum systems through AWS Bracket~\cite{GonzalesCloud2021}. On top of that, the race among quantum hardware makers for more qubit count such as IBM's annoucement of a targeted 1000-qubit quantum system~\cite{RielQuantum2021} is directed at the goal of realizing general-purpose quantum systems capable of solving problems intractable for today's conventional classical systems.
	
	Studying the implementation of these systems, a common challenge among them is the limited connectivity between the physical qubits. This is due to the fact that, with an increasing degree of connectivity among qubits, the possibility of frequency collision among qubits and degeneracies among transition frequencies increases substantially~\cite{IBM2021}. To circumvent these problems, quantum hardware makers have devised layouts that are sparsely connected. In this vein, common layout schemes are octagonal as followed by Rigetti~\cite{RigettiAspen2023} or linear as well as hexagonal as used by IBM~\cite{IBM2021}.
	
	Eventually, the (limited) interactions of these layouts result in different \emph{superconducting quantum computing architectures}, which are described as a graph known as the \emph{coupling graph}. In these graphs, each qubit is represented by a vertex and each edge represents a possible interaction between two qubits. Qubits that are connected (i.e, that are adjacent on the graph) can perform 2-qubit gates. However, physical limitations on the connectivity of qubits mean that the degree of the qubit interaction is low, which restricts the ability of qubits to interact with each other.
	
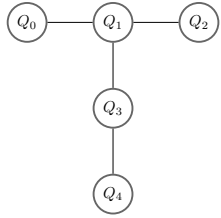
\begin{figure}
		\centering
		\scalebox{0.6}{
			\begin{tikzpicture}[
				roundnode/.style={circle, draw=black!60, very thick, minimum size=7mm}
				]
				\node[roundnode]    (0) {$Q_0$};
				\node[roundnode]    (1) [right=of 0] {$Q_1$};
				\node[roundnode]    (2) [right=of 1] {$Q_2$};
				\node[roundnode]    (3) [below=of 1] {$Q_3$};
				\node[roundnode]    (4) [below=of 3] {$Q_4$};
				\draw[-] (0.east) -- (1.west);
				\draw[-] (1.east) -- (2.west);
				\draw[-] (1.south) -- (3.north);
				\draw[-] (3.south) -- (4.north);
				
			\end{tikzpicture}
		}
		\caption{Coupling graph for the \textit{ibmq\_lima} quantum device}
		\label{fig:ibm_lima_coupling_graph}
	\end{figure}

	\begin{example}
		\autoref{fig:ibm_lima_coupling_graph} sketches the coupling graph of the \textit{ibmq\_lima} architecture---one of the IBM Quantum Falcon Processors~\cite{IBMCompute2021}. As can be seen, this allows qubit interactions and, hence, operations between, e.g., qubits~$Q_0$ and~$Q_1$ but not, e.g., between qubits~$Q_0$ and~$Q_2$.
	\end{example}
	
	Equipped with the knowledge of limited connectivity in the layout, \emph{coupling constraints} can be defined as a set of rules that determine which qubit(s) can interact with a particular qubit---directly imposing restrictions on the permissible two-qubit gate interactions.
	
	Apart from the coupling constraints in the layout, these architectures also do not support arbitrary gates that are possible to be executed from the software stack. For instance, current IBM devices support Pauli X, Y, Z gates, the Phase gates, the square-root of Phase gates, CNOT, controlled U gates, and \emph{Echoed Cross-Resonance} (ECR) gates with new gate support being added over time. The gates that are supported directly on a quantum architecture are known as elementary or basis gates.
	
	\subsection{Satisfying Coupling Constraints}
	
	Having superconducting architectures with their corresponding coupling constraints as reviewed above obviously limits the arbitary execution of quantum circuits. In fact, these architectures only allow interactions between qubits connected to each other. This limitation can be overcome by clever initial logical to physical qubit mapping and insertion of SWAP gates, which we will illustrate in the next example.
	
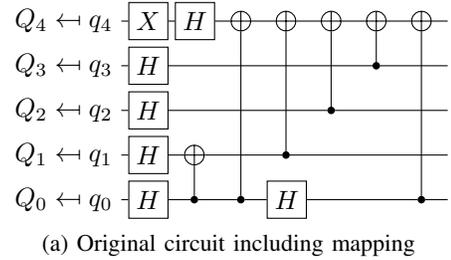
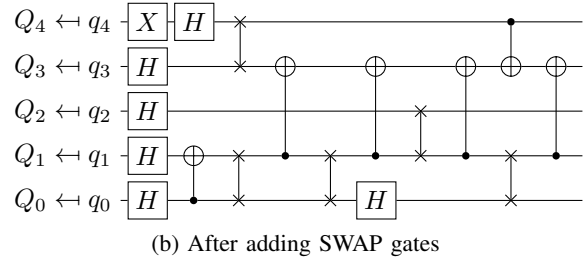
\begin{figure}
		\centering
		\begin{subfigure}{0.45\textwidth}
			\centering
			\begin{tikzpicture}
				\begin{yquant}
					qubit {$Q_4 \mapsfrom q_4$} q;    	
					qubit {$Q_3 \mapsfrom q_3$} q[+1];
					qubit {$Q_2 \mapsfrom q_2$} q[+1];
					qubit {$Q_1 \mapsfrom q_1$} q[+1];
					qubit {$Q_0 \mapsfrom q_0$} q[+1];
					
					x q[0];
					h q[1];
					h q[2];
					h q[3];
					h q[4];
					
					h q[0];
					cnot q[3] | q[4];
					
					cnot q[0] | q[4];
					h q[4];
					cnot q[0] | q[3];
					cnot q[0] | q[2];
					cnot q[0] | q[1];
					cnot q[0] | q[4];
				\end{yquant} 
			\end{tikzpicture}
			\caption{Original circuit including mapping}
			\label{fig:example_quantum_circuit_original}
		\end{subfigure}
		\par\bigskip
		\begin{subfigure}{0.45\textwidth}
			\centering
			\begin{tikzpicture}
				\begin{yquant}
					qubit {$Q_4 \mapsfrom q_4$} q;    	
					qubit {$Q_3 \mapsfrom q_3$} q[+1];
					qubit {$Q_2 \mapsfrom q_2$} q[+1];
					qubit {$Q_1 \mapsfrom q_1$} q[+1];
					qubit {$Q_0 \mapsfrom q_0$} q[+1];
					
					x q[0];
					h q[1];
					h q[2];
					h q[3];
					h q[4];
					
					h q[0];
					cnot q[3] | q[4];
					
					swap (q[0], q[1]);
					swap (q[3], q[4]);
					
					cnot q[1] | q[3];
					
					swap (q[3],q[4]);
					
					h q[4];
					cnot q[1] | q[3];
					
					swap (q[2], q[3]);
					
					cnot q[1] | q[3];
					
					swap (q[3], q[4]);
					cnot q[1] | q[0];
					
					cnot q[1] | q[3];
				\end{yquant} 
			\end{tikzpicture}
			\caption{After adding SWAP gates}
			\label{fig:example_quantum_circuit_swap_inserted}
		\end{subfigure}
		\caption{Satisfying coupling constraints}
		\label{fig:example_quantum_circuit}\vspace{-.4cm}
	\end{figure}

	\begin{example}
		Consider the quantum circuit shown in \autoref{fig:example_quantum_circuit_original}. Executing this quantum circuit on the architecture represented by the coupling graph shown in \autoref{fig:ibm_lima_coupling_graph} requires all qubits~$q_0, q_1, \dots$ of the circuit to be mapped to the qubits~$Q_0, Q_1, \dots$ of the architecture. A possible mapping is indicated on the left-hand side of \autoref{fig:example_quantum_circuit_original} through the notation ${Q_i} \mapsfrom q_j$. While this allows, e.g., for the execution of the first CNOT gate (requiring an interaction between~$Q_0$ and~$Q_1$), the second gate (requiring an interaction between~$Q_0$ and~$Q_4$) can \emph{not} be executed on that architecture (since, according to \autoref{fig:ibm_lima_coupling_graph}, the architecture does not allow for this interaction). Hence, SWAP gates have to be added as shown, e.g., in \autoref{fig:example_quantum_circuit_swap_inserted}, which ``move'' the qubits to positions within the architecture that allow for the desired interactions. Obviously, these SWAP gates increase the overall gate count and, hence, harms the fidelity of the circuit and makes the execution more prone to errors/noise.
	\end{example}

	In the past, numerous compilers (also known as transpilers) and corresponding methods have been proposed to address this problem---including heuristic~\cite{LiTackling2019,Cowtan2019,zulehner2018efficient,zulehner2019compiling,wille2023mqt} as well as exact solutions~\cite{WilleMapping2019,Tan2020}. However, they have to cope with the constraints enforced by the provided quantum architecture. In this work, we further consider a complementary approach that, additionally, considers changing the architecture itself. 
	
	\section{Design and Realization\\of Alternative Architectures}
	\label{sec:design_and_realization_of_alternative_architectures}
	
	Instead of taking a given architecture as an invariant, an additional and complementary way of reducing the number of correspondingly needed SWAP gates is to design and realize an alternative (application-specific) architecture. This architecture obviously still has to satisfy physical constraints to be realizable but may be designed in a fashion that is more suited for the needed interactions of a given quantum circuit (or a set thereof). In this section, we first briefly review the underlying idea (as it has also been employed in some initial related works) and discuss the current drawbacks in this state-of-the-art. Based on that, we then describe the main motivations and contributions of the framework proposed in this work, which aim at addressing these drawbacks.

	\subsection{Application-specific Quantum Hardware Architecture}
	Reducing the gate count while adhering to the physical constraints of the quantum architecture are the main objectives of today's compilers/transpilers. They map a given application in terms of a quantum circuit to the targeted quantum architecture. These quantum architectures, in turn, are usually not designed and realized with any particular application in mind.
	
	However, there is a certain degree of freedom in which alternative architectures can be realized. This freedom can be used in order to get an architecture which is more suited for a given (set of) quantum algorithms. Alternative architectures optimized for an application can reduce the number of SWAP gates and, hence, increase the fidelity of a circuit substantially.
	
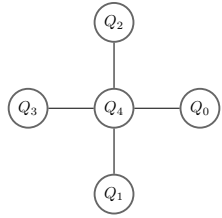
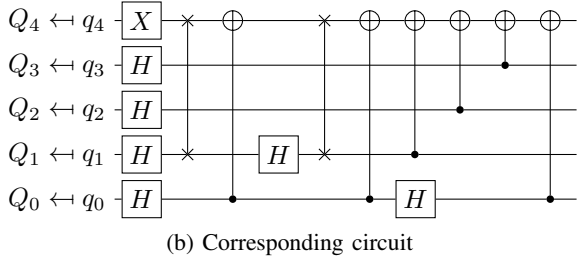
\begin{figure}
		\centering
		\begin{subfigure}{0.45\textwidth}
			\centering
			\scalebox{0.6}{
				\begin{tikzpicture}[
					roundnode/.style={circle, draw=black!60, very thick, minimum size=7mm}
					]
					\node[roundnode]    (4) {$Q_4$};
					\node[roundnode]    (3) [left=of 4] {$Q_3$};
					\node[roundnode]    (0) [right=of 4] {$Q_0$};
					\node[roundnode]    (2) [above=of 4] {$Q_2$};
					\node[roundnode]    (1) [below=of 4] {$Q_1$};
					\draw[-] (4.north) -- (2.south);
					\draw[-] (4.south) -- (1.north);
					\draw[-] (4.east) -- (0.west);
					\draw[-] (4.west) -- (3.east);        
				\end{tikzpicture}
			}
			\caption{Alternative architecture}
			\label{fig:application_driven_coupling_graph}
		\end{subfigure}
		\par\bigskip
		\begin{subfigure}{0.45\textwidth}
			\centering
			\begin{tikzpicture}
				\begin{yquant}
					qubit {$Q_4 \mapsfrom q_4$} q;    	
					qubit {$Q_3 \mapsfrom q_3$} q[+1];
					qubit {$Q_2 \mapsfrom q_2$} q[+1];
					qubit {$Q_1 \mapsfrom q_1$} q[+1];
					qubit {$Q_0 \mapsfrom q_0$} q[+1];
					
					x q[0];
					h q[1];
					h q[2];
					h q[3];
					h q[4];
					
					swap (q[0], q[3]);
					
					cnot q[0] | q[4];
					
					h q[3];
					
					swap (q[0], q[3]);
					
					cnot q[0] | q[4];
					h q[4];
					
					cnot q[0] | q[3];
					cnot q[0] | q[2];
					cnot q[0] | q[1];
					cnot q[0] | q[4];		
				\end{yquant}
			\end{tikzpicture}
			\caption{Corresponding circuit}
			\label{fig:application_driven_quantum_circuit_swap_inserted}
		\end{subfigure}
		\caption{Application-specific quantum hardware architecture}
		\label{fig:application_driven_quantum_hardware_architecture}
	\end{figure}

	\begin{example}
		\autoref{fig:application_driven_coupling_graph} shows an alternative application-specific architecture for the quantum circuit shown in \autoref{fig:example_quantum_circuit_original}. The quantum circuit mapped to this alternative architecture is shown in \autoref{fig:application_driven_quantum_circuit_swap_inserted}. Using the alternative architecture, only two SWAP gates are needed compared to five SWAP gates in the quantum circuit mapped to the original architecture shown in \autoref{fig:example_quantum_circuit_swap_inserted}. Consequently, the quantum circuit mapped to the alternative architecture may achieve higher circuit fidelity as lesser number of SWAP gates used.
	\end{example}
	
	To this end,~\cite{Deb2020,Lin2022} proposed several schemes to determine alternative architectures that allow for more efficient mapping of quantum circuits to them. However, although these approaches provided a pathway for the identification of alternative architectures, the question still remained how to realize those architectures using actual hardware. In this regard,~\cite{Li2019} goes one step further than~\cite{Deb2020,Lin2022} by proposing an automatic design flow to generate a \emph{high-level quantum architecture} for different quantum algorithms. In spite of that,~\cite{Li2019} did not map the \emph{high-level quantum architecture} to physical design tools to verify the validity of the architecture.
	
	Hence, although these application-specific quantum hardware architectures have their advantages, there is still a question on whether these architectures are valid architectures that are physically realizable. This open question will be the focus of the next section.
	
	\subsection{Physical Design of Quantum Hardware Architecture}
	
	In this section, we review the process used to generate a physical design of an (alternative) quantum hardware architecture. The process can be split into distinct phases. First, an initial physical layout is generated. Afterwards, an iterative process of analyzing and finetuning the design is conducted until a working result is obtained. Each phase is elaborated further in the following.
	
	The initial physical layout is generated by realizing the components in the layout. As a key component, qubits are constructed from an element that introduces non-linearity to the qubits. Non-linearity is needed to encode and manipulate quantum information in the quantum system composed of these qubits~\cite{Blais2021}. In this regard, Josephson junctions~\cite{Josephson1962} are the most frequently used non-linear element. A detailed description of the physics behind Josephson junctions can be found in~\cite{Goran2017}. Using the non-linearity introduced by Josephson junctions, there are many variants of qubits such as transmon, flux qubit~\cite{Mooij1999}, and fluxonium~\cite{Manucharyan2009} that can be used with transmon qubits being the most ubiquitous variant~\cite{Koch2007}.
	
	The initial physical layout generation is continued through the construction of connections between qubits using resonators. Resonators play the role of controlling and measuring the qubits~\cite{RasmussenSuperconducting2021}. The length of a resonator determines its frequency and its ability to couple with qubits. To this end, the length of a resonator can be estimated by computing the wavelength of the coaxial traveling modes given a target frequency~\cite{Gao2021}. After the qubits and resonators are in place, the readouts/controls are added to read signals to and from the chip for which resonators are used. Additionally, capacitors are added to make the system more sensitive to the changes in the qubit and to control the electric field distribution in the system~\cite{Krantz2019}. Finally, the connections between qubits and capacitors and capacitors and readouts/controls are realized. This concludes the initial physical layout phase.
	
\begin{figure}
		\centering
		\includegraphics[scale=0.5]{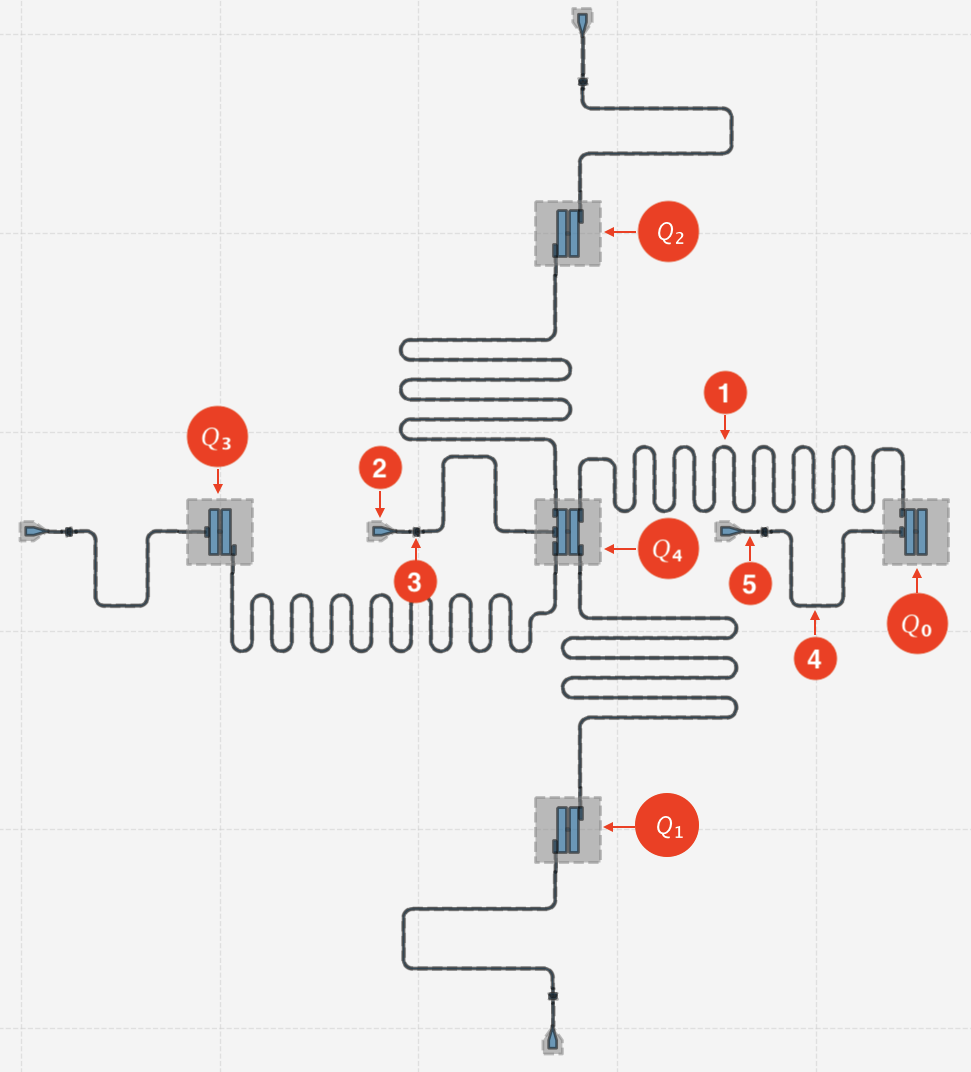}
		\caption{Physical layout of high-level architecture}
		\label{fig:physical_layout_architecture_qiskit_metal}\vspace{-.4cm}
	\end{figure}

	\begin{example}
		Consider the application-specific quantum hardware architecture shown in \autoref{fig:application_driven_coupling_graph}. Realizing this architecture will require an initial physical layout as shown in \autoref{fig:physical_layout_architecture_qiskit_metal}. As indicated by the annotations in the image, qubit~$Q_4$ is placed at the center of the layout, while qubits~$Q_0$\dots$Q_3$ are placed around qubit~$Q_4$. After placing the qubits, the qubit-to-qubit connections are realized using resonators \textcircled{\scalebox{0.8}{$1$}}. Following this, the readout/control \textcircled{\scalebox{0.8}{$2$}} and capacitors \textcircled{\scalebox{0.8}{$3$}} are placed. Finally, to conclude the initial physical layout, the connection between qubits and capacitors \textcircled{\scalebox{0.8}{$4$}} as well as capacitors and readouts/controls \textcircled{\scalebox{0.8}{$5$}} are constructed.
	\end{example}
	
	Having a first placement and corresponding connections, the first design of the physical realization of the alternative architecture is obtained. However, to make it work, several further parameters need to be determined. For instance, the target parameters for qubits are composed of its frequency, anharmonicity, ratio~($E_j/E_c$) of the Josephson junction's inductive energy~$E_j$ and capacitive energy~$E_c$, relaxation time~$T_1$, and dephasing time~$T_2$. Similarly, the resonator frequency is also a parameter that needs to be optimized to realize a workable quantum circuit. These parameters are in no way exhaustive, as there are various elements interacting in a quantum chip design. More in-depth discussions of the these and other metrics can be found in~\cite{Gao2021,Krantz2019}.
	
	Having all these parameters set, a first complete design of an application-specific architecture is obtained. However, having such a design does not mean that it actually works. In fact, generating a \emph{valid} design requires a clever orchestration and refinement of the constructed placements, connections, and all parameters. Hence, obtained designs are usually subject to detailed analysis and revisions until a solution is obtained that actually works and can be fabricated. To this end, simulation tools such as Ansys HFSS, Comsol Multiphysics, and ElmerFEM can be used to extract the results needed for analysis. In this context, Qiskit Metal~\cite{Minev2021} is a tool that eases the simulation process by automatically rendering the physical layout in the simulation software and seamlessly extracting the required results without much manual intervention.
	
	Using these results, the analysis techniques quantize the Hamiltonian of a superconducting quantum circuit system. A Hamiltonian is a function that is equal to the total kinetic and potential energy of a system, which also corresponds to the total energy of the system~\cite{QiskitTextbook}. Some examples of recent quantization techniques are the \emph{Lumped Oscillator Model}~(LOM,~\cite{MinevLOM2021}) and the \emph{Energy Participation Ratio}~(EPR,~\cite{MinevEPR2021}). A more detailed exposition on the similarities and differences between these two methods can be found in~\cite{Yuan2022}. From a practical point of view, Qiskit Metal~\cite{Minev2021} can be used again to run some of the analysis techniques supported in the tool, such as LOM and EPR.
	
	Given that the analysis step is concluded, the design is fine-tuned by adjusting the geometries of the components in the physical layout to move the value of the parameters closer to the acceptable target parameters.
	
	\subsection{Drawbacks and Contributions}
	
	The methodology presented above provides a promising procedure to create alternative superconducting quantum architectures that perform better and more reliably for the given, dedicated applications. However, a major drawback of the current state of the art is that tool support for the single tasks is hardly available or, if available, hardly integrated thus far. This leaves the designer rather alone with a design task which, additionally, requires expertise from several domains (such as architecture design and physical design). More precisely, based on the current state of the art,
	\begin{enumerate}
		\item a designer needs to manually figure out which architecture executes their application with the best possible circuit fidelity and least circuit depth,
		\item a designer needs to manually map the high-level architecture to a physical layout, and 
		\item a designer needs to manually optimize the geometries of his/her design until he/she determines a working solution. 
	\end{enumerate}
	All these steps are time-consuming, error-prone, and require corresponding expertise as well as background.
	
	Motivated by this, in this artifact paper, a tool framework is proposed that (1)~aims at providing a foundation to integrate the corresponding tools together, (2)~provides automated support to ``bridge'' architecture design and physical design, and, by this, (3)~addresses the drawbacks mentioned above.
	More precisely, the proposed framework provides
	
	\begin{enumerate}
		\item a well-defined framework that streamlines the manual stages of designing an application-specific quantum hardware architecture into a unified automated workflow,
		\item a reference implementation
		that can be executed in a ``push-button'' fashion---allowing even non-experienced designers to automatically create application-specific superconducting quantum architectures, and
		\item a modular structure that can effortlessly be extended to incorporate diverse alternative approaches and methods for generating and optimizing the hardware architecture within the framework (without the need to completely overhaul the framework and/or start from scratch once a new approach is developed).
	\end{enumerate}

	In the next section, the proposed framework as well as how to use it and how to extend it are described in more detail.
	
	\section{Proposed Framework}
	\label{sec:proposed_framework}
	
	The application-specific quantum hardware architecture framework proposed in this work encompasses the high-level generation and physical layout of the architecture. The framework is built on the basis of a modular and easily extensible implementation to enable ease of use and integration of improvements in the future. In this section, we will provide an overview of the framework and its implementation as well as a description how to use and how to extend the framework. The framework itself is publicly available at
	\href{https://github.com/cda-tum/dasqa}{https://github.com/cda-tum/dasqa}.
	
	\subsection{Overview of the Framework}
	
\begin{figure*}[t]
		\centering
		\resizebox{\columnwidth*2}{!}{
			\begin{tikzpicture}[
				rectanglenode/.style={rectangle, draw=black!60, fill=black!5, very thick}
				]
				\node [] (quantum_circuit_image) {
					\begin{tikzpicture}[]
						\begin{yquant}
							qubit {$Q_4 \mapsfrom q_4$} q;    	
							qubit {$Q_3 \mapsfrom q_3$} q[+1];
							qubit {$Q_2 \mapsfrom q_2$} q[+1];
							qubit {$Q_1 \mapsfrom q_1$} q[+1];
							qubit {$Q_0 \mapsfrom q_0$} q[+1];
							
							x q[0];
							h q[1];
							h q[2];
							h q[3];
							h q[4];
							
							h q[0];
							cnot q[3] | q[4];
							
							cnot q[0] | q[4];
							h q[4];
							cnot q[0] | q[3];
							cnot q[0] | q[2];
							cnot q[0] | q[1];
							cnot q[0] | q[4];
						\end{yquant}        
					\end{tikzpicture}
				};
				\node [] (quantum_circuit_label) [below=1mm of quantum_circuit_image] {Quantum circuit};
				\node [rectanglenode] (architecture_generator) [right=of quantum_circuit_image] {Architecture generator};
				\node (architecture_generator_output) [right=of architecture_generator]{
					\scalebox{0.6}{
						\begin{tikzpicture}[
							roundnode/.style={circle, draw=black!60, very thick, minimum size=7mm}
							]
							\node[roundnode]    (4) {$Q_4$};
							\node[roundnode]    (3) [left=of 4] {$Q_3$};
							\node[roundnode]    (0) [right=of 4] {$Q_0$};
							\node[roundnode]    (2) [above=of 4] {$Q_2$};
							\node[roundnode]    (1) [below=of 4] {$Q_1$};
							\draw[-] (4.north) -- (2.south);
							\draw[-] (4.south) -- (1.north);
							\draw[-] (4.east) -- (0.west);
							\draw[-] (4.west) -- (3.east);        
						\end{tikzpicture}
					}
				};
				\node[] (architecture_generator_output_label) [below=1mm of architecture_generator_output] {High-level architecture};
				\node [rectanglenode] (physical_layout_mapper) [right=of architecture_generator_output] {Physical layout mapper};
				\node [rectanglenode] (optimizer) [right=of physical_layout_mapper] {Optimizer};
				\node [] (physical_layout_output) [right=of optimizer]{
					\includegraphics[scale=0.2]{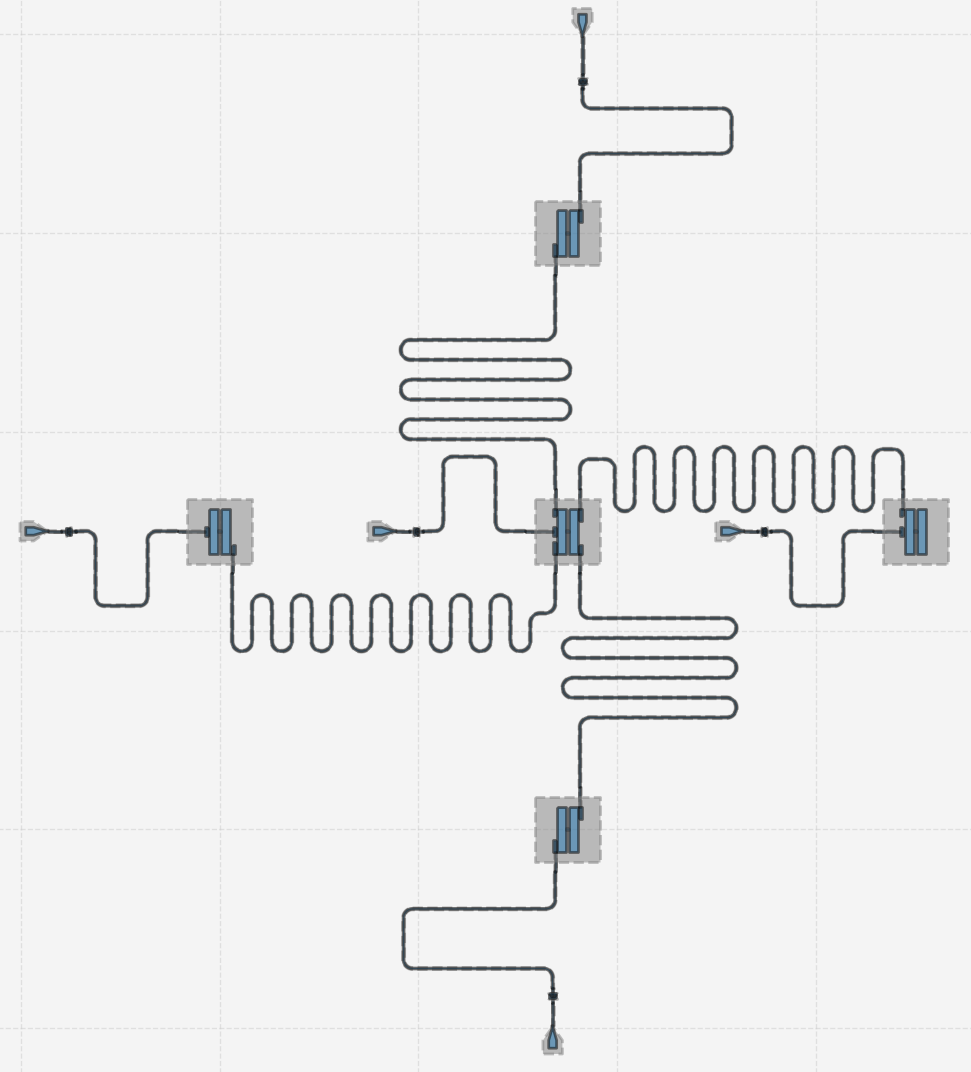}
				};
				\node[align=center] (physical_layout_output_label) [below=1mm of physical_layout_output] {Physical layout\\of architecture};
				\draw[->, shorten >= 5pt, shorten <= 5pt] (quantum_circuit_image.east) -- (architecture_generator.west);
				\draw[->, shorten >= 5pt, shorten <= 5pt] (architecture_generator.east) -- (architecture_generator_output.west);
				\draw[->, shorten >= 5pt, shorten <= 5pt] (architecture_generator_output.east) -- (physical_layout_mapper.west);
				\draw[->, shorten >= 5pt, shorten <= 5pt] (physical_layout_mapper.east) -- (optimizer.west);
				\draw[->, shorten >= 5pt, shorten <= 5pt] (optimizer.east) -- (physical_layout_output.west); 
				
				\node [fit={(quantum_circuit_image) (architecture_generator) (architecture_generator_output) (physical_layout_mapper)}, inner sep=10pt] (architecture_generator_bracket) {};
				
				\coordinate (halfway_right_architecture_generator_bracket) at (physical_layout_mapper.center |- architecture_generator_bracket.north);
				
				\draw [decorate, decoration={brace, amplitude=10pt, raise=12pt}] (architecture_generator_bracket.north west) -- (halfway_right_architecture_generator_bracket) node [midway, above=24pt] {\textbf{Phase 1: Architecture generation}};
				
				\node [fit={(physical_layout_mapper) (optimizer) (physical_layout_output)}, inner sep=10pt] (physical_design) {};
				
				\coordinate (halfway_left_physical_design) at (physical_layout_mapper.center |- physical_design.north);
				
				\draw [decorate, decoration={brace, amplitude=10pt}] (halfway_left_physical_design) -- (physical_design.north east) node [midway, above=12pt] {\textbf{Phase 2: Physical design}};
			\end{tikzpicture}
		}
		\caption{Design flow of framework}
		\label{fig:design_flow_of_framework}
	\end{figure*}
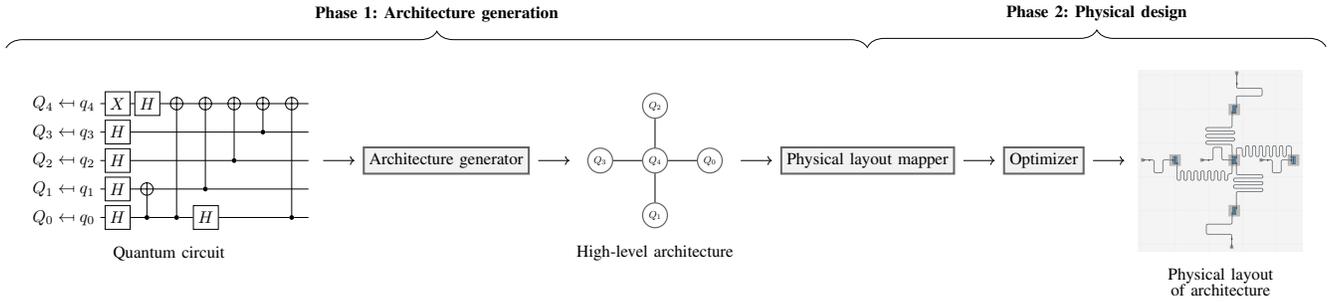

	The framework proposed in this work provides an integrated solution that can take arbitrary solutions for the architecture generation and optimization steps as well as execute the relevant steps in \autoref{sec:design_and_realization_of_alternative_architectures} in an automated fashion. The framework weaves together these intermediate steps to take a quantum application (i.e, a quantum circuit) as input and to generate a physical layout that is optimized for the given quantum circuit. In this section, we will describe the structure of the framework as shown in \autoref{fig:design_flow_of_framework} and how it overcomes the drawbacks mentioned in the previous section.
	
	First, the framework provides an interface through the architecture generator to execute an architecture generation algorithm defined by the user (such as those provided in \cite{Li2019} and discussed above). The architecture generator is intended to generate the high-level architecture of a quantum architecture that is optimized for a quantum application. Through this step, the user does not need to manually figure out which architecture is the most optimized for their quantum application.
	
	Next, the physical layout mapper acts as an intermediary between the architecture generation and the physical design phase. The mapper provides an approach to automate the mapping process of the high-level architecture to the initial physical layout, which has been mostly done manually so far.
	
	Finally, the optimizer provides the possibility to execute different algorithms that define the geometries that are appropriate for a component given the target parameter that needs to be achieved. The inclusion of an automated optimizer saves the user from manually optimizing the geometries to hit the target parameters, saving time and possible errors. 
	
	\subsection{Implementation of the Framework}
	
	The framework is implemented in a generic manner using Python for ease of use and extensibility.
	The architecture generation and optimization steps as shown in \autoref{fig:design_flow_of_framework} are implemented as abstract methods. Abstract methods are methods without implementation that need to be overridden by subclasses with their own implementation. The architecture generation and physical layout mapping steps are implemented as abstract methods in the framework to allow subclasses to plug-in their own architecture generation and optimization algorithm. The physical layout mapping step is an implementation provided in the framework based on Qiskit Metal~\cite{Minev2021} that does not need overriding. 
	
	More precisely, the architecture generator's abstract method is defined as:
	
\begin{lstlisting}[language=python]
@abstractmethod
def generate_architecture(
		self, qc: QuantumCircuit, config: dict
	) -> tuple[np.ndarray, np.ndarray]:
	pass
\end{lstlisting}
	
	\noindent The quantum application (i.e, a quantum circuit) in the form of a Qiskit \lstinline[basicstyle=\ttfamily\normalsize]|QuantumCircuit| object is passed to the subclasses implementing the method as the argument \lstinline[basicstyle=\ttfamily\normalsize]|qc|. In addition, the configuration dictionary read from the user-provided configuration file is passed as argument \lstinline[basicstyle=\ttfamily\normalsize]|config| which can then be used in the architecture generation algorithm. The subclasses implementing the method needs to return a Numpy 2D matrix containing the qubit layout and a Numpy 1D array containing the qubit frequencies as denoted by the return type hints \lstinline[basicstyle=\ttfamily\normalsize]|np.ndarray| in the abstract method definition. The qubit layout needs to have the location of the qubits defined as elements of the matrix with $-1$ on positions without qubits. 
	
	\begin{example}
		Consider the high-level architecture in \autoref{fig:design_flow_of_framework}.  The qubit layout needs to be represented by a 2D matrix as:
		\[
		\begin{bmatrix}
			$-1$ & $2$ & $-1$\\
			$3$ & $4$ & $0$\\
			$-1$ & $1$ & $-1$
		\end{bmatrix}
		\]
		Matrix element $0$ indicates the position of qubit $Q_0$ in the layout and, following this convention, the location of other qubits (i.e, $Q_1$, $Q_2$,\dots) is specified in the matrix. Matrix element $-1$ indicates no qubit at that position in the layout. Correspondingly, $\begin{bmatrix}$5.06$, $5.24$, $5.08$, $5.27$, $5.17$\end{bmatrix}$ is an example of a 1D array containing the qubit frequencies where each element starting from index $0$ corresponds to the frequency of the respective qubit in the qubit layout (e.g., $5.06$ is the frequency of qubit $Q_0$).
	\end{example}
	
	In a similar fashion, the optimizer function's abstract method is defined as:
	
\begin{lstlisting}[language=python]
@abstractmethod
def optimize_layout(
		self, canvas: CanvasBase,
		qubit_frequencies: np.ndarray, config: dict,
	):
	pass
\end{lstlisting}
	
	\noindent The \lstinline[basicstyle=\ttfamily\normalsize]|canvas| argument is a \lstinline[basicstyle=\ttfamily\normalsize]|CanvasBase| object encapsulating the Qiskit Metal design object built in the mapper step. The geometry of a component in the Qiskit Metal design can be updated in the optimization process by calling \lstinline[language=python, basicstyle=\ttfamily\normalsize]|canvas.update_component(component_name,option_name, option_value)|. Besides the \lstinline[basicstyle=\ttfamily\normalsize]|canvas|%
	object, the qubit frequencies returned in the architecture generation stage and the configuration dictionary that contains the user defined target parameters is passed to the optimization algorithm. The configuration dictionary is the same dictionary passed in the architecture generation step. Hence, there is no need for the user to define a separate configuration file.
	
	\begin{example} Given the target parameters (e.g, $E_j/E_c$ ratio) are defined in the configuration file, the information in the configuration file and the qubit frequencies from the architecture generation step will be passed to the subclass implementing the  \lstinline[basicstyle=\ttfamily\normalsize]|optimize_layout| method. Using these information, the optimizer can find optimal geometries of the components and update the geometries of the components. For instance, the optimizer can update the pad gap of qubit $Q_0$ to $10$um by calling \lstinline[language=python,basicstyle=\ttfamily\normalsize]|canvas.update_component("Q_0","pad_gap","10um")|. 
	\end{example}
	
	Equipped with the knowledge of the framework's implementation, the next sections cover how to use and how to extend the framework. 
	
	\subsection{Using the Framework}
	
	To illustrate the applicability of the proposed framework, a subclass containing a reference implementation is available in the Python file \emph{src/concrete\_design\_flow1.py}. An examplary implementation of architecture generator is implemented based on \cite{Li2019} which can be found in \emph{src/architecture\_generator1} module. This implementation is called in the \lstinline[basicstyle=\ttfamily\normalsize]|generate_architecture| method in the subclass. 
	
	\textls[-32]{Next, the optimizer is implemented by the \lstinline[basicstyle=\ttfamily\normalsize]|optimize_layout| method.} In the reference implementation, we have implemented a statistical model using polynomial regression as the algorithm to yield the pad gap and pad height of a qubit given a target qubit frequency. The statistical model is trained on pre-collected simulation data and acts as a stand-in for simulation softwares. More details on the statistical model and optimization module can be found in the \emph{src/optimal\_geometry\_finder} and \emph{src/optimizer} modules respectively.
	
	With these instantiations of the abstract methods defined, the framework automatically takes the quantum application (i.e, quantum circuit) as input, executes the architecture generation algorithm, and maps the high-level architecture to an initial layout using Qiskit Metal. This initial physical layout is then optimized by executing the optimization algorithm, and displaying it to the user in a GUI. Interested readers can follow the installation instructions in our repository and try the reference implementation by executing a CLI command
	
\begin{lstlisting}[language=bash, label={lst:cli}]
dasqa --file-path ./src/tests/test_circuit/circuit1.qasm --config-file-path ./src/tests/test_config/config.yml
\end{lstlisting}
	
	\noindent The arguments of the CLI command are the path of the quantum circuit and configuration file passed as \lstinline|--file-path| and \lstinline|--config-file-path|, respectively.
	
	\subsection{Extending the Framework}
	
	As previously discussed, obvious extensions of the framework are 
	different implementations for architecture generation and the optimization step. 
	In this context, each module in the reference implementation has abstract classes defined which allows different implementations to be realized without the need to replace the whole module. This allows for a fast implementation of different algorithms to ease experimentation.
	However, the framework is extensible beyond these steps to further ease experimentation. 
	As an example, the qubit layout subclass in the physical layout mapper inherits from the \emph{QubitBase} abstract class as:
\begin{lstlisting}[language=python, label={lst:qubit}]
class QubitBase(ABC):
	@abstractmethod
	def generate_qubit_layout(self):
		pass
\end{lstlisting}
	
	\noindent The qubit layout subclass then defines how the qubits are positioned on the physical layout. For a complete list of possible abstract classes that can be overriden in each module, we point the reader to refer to our open-sourced code repository.
	
	\section{Conclusion}
	\label{sec:conclusion}
	
	Superconducting quantum hardware architectures are usually designed with physical constraints in mind. These architectures can be improved by tailoring it to specific applications. To this end, this artifact paper proposed a framework and a reference implementation for the design and realization of alternative, i.e., application-specific, superconducting quantum architectures. The resulting framework covers the corresponding design flow in an integrated and automated as well as modular fashion---allowing designers to realize the desired design in a push-button manner as well as to easily experiment with different design choices. The framework with a reference implementation is available as part of the \emph{Munich Quantum Toolkit}~(MQT) at \href{https://github.com/cda-tum/dasqa}{https://github.com/cda-tum/dasqa} under an \mbox{open-source} license. By this, we hope to trigger more interest on the area of designing application-specific quantum hardware.
	
\section*{Acknowledgments}
Funded by the European Union under Horizon Europe Programme - Grant Agreement 101080086 — NeQST. Views and opinions expressed are however those of the author(s) only and do not necessarily reflect those of the European Union or the European Commission. Neither the European Union nor the granting authority can be held responsible for them.
	
	\printbibliography
\end{document}